\documentclass[acmsmall]{acmart}

\PassOptionsToPackage{table,xcdraw}{xcolor} 
\usepackage[flushleft]{threeparttable}
\usepackage{multirow}
\usepackage{rotating}
\usepackage{colortbl}
\usepackage{subcaption}
\usepackage[normalem]{ulem}
\usepackage{csquotes}
\usepackage{tabularx}

\AtBeginDocument{%
  \providecommand\BibTeX{{%
    \normalfont B\kern-0.5em{\scshape i\kern-0.25em b}\kern-0.8em\TeX}}}

\setcopyright{rightsretained}
\acmJournal{PACMHCI}
\acmYear{2024} \acmVolume{8} \acmNumber{CSCW2} \acmArticle{370} \acmMonth{11} \acmPrice{15.00} \acmDOI{10.1145/3686909}

\received{July 2023}
\received[revised]{April 2024}
\received[accepted]{July 2024}

\begin{document}

\title[``Sharing, Not Showing Off'': How BeReal Approaches \\ Authentic Self-Presentation on Social Media Through Its Design]{``Sharing, Not Showing Off'': How BeReal Approaches Authentic Self-Presentation on Social Media Through Its Design}

\author{JaeWon Kim}
\email{jaewonk@uw.edu}
\orcid{0000-0002-9319-2704}
\affiliation{%
  \institution{University of Washington}
  \state{WA}
  \country{USA}
}

\author{Robert Wolfe}
\email{rwolfe3@uw.edu}
\orcid{0000-0001-7133-695X}
\affiliation{%
  \institution{University of Washington}
  \state{WA}
  \country{USA}
}

\author{Ishita Chordia}
\email{ichordia@uw.edu}
\orcid{0000-0002-9319-2704}
\affiliation{%
  \institution{University of Washington}
  \state{WA}
  \country{USA}
}

\author{Katie Davis}
\email{kdavis78@uw.edu}
\orcid{0000-0001-8794-8651}
\affiliation{%
  \institution{University of Washington}
  \state{WA}
  \country{USA}
}

\author{Alexis Hiniker}
\email{alexisr@uw.edu}
\orcid{0000-0003-1607-0778}
\affiliation{%
  \institution{University of Washington}
  \state{WA}
  \country{USA}
}

\renewcommand{\shortauthors}{JaeWon Kim et al.}


\begin{abstract}
    Adolescents are particularly vulnerable to the pressures created by social media, such as heightened self-consciousness and the need for extensive self-presentation. In this study, we investigate how \textit{BeReal}, a social media platform designed to counter some of these pressures, influences adolescents' self-presentation behaviors. We interviewed 29 users aged 13-18 to understand their experiences with BeReal. We found that BeReal's design focuses on spontaneous sharing, including randomly timed daily notifications and reciprocal posting, discourages staged posts, encourages careful curation of the audience, and reduces pressure on self-presentation. The space created by BeReal offers benefits such as validating an unfiltered life and reframing social comparison, but its approach to self-presentation is sometimes perceived as limited or unappealing and, at times, even toxic. Drawing on this empirical data, we propose design guidelines for platforms that support authentic self-presentation while fostering reciprocity and expanding beyond spontaneous photo-sharing. These guidelines aim to enable users to portray themselves more comprehensively and accurately, ultimately supporting teens' developmental needs, particularly in building authentic relationships.
\end{abstract}



\maketitle

\section{Introduction}

The prevalence of social isolation has increased dramatically in recent decades. The number of people reporting no confidants for discussing important matters has tripled, while the average person's social network has shrunk by one-third~\cite{mcpherson2006social}. By 2004, nearly 25\% of Americans reported having no confidant, up from 10\% in 1985~\cite{mcpherson2006social}. This decline in social connections is concerning, as relationships significantly impact physical and mental health, longevity, and happiness~\cite{cacioppo2014social, house1988social, cohen2004social}. The 75-year Study of Adult Development found that strong relationships are key to happiness and health~\cite{waldinger2016long}. A 2010 meta-analysis of over 300,000 participants revealed that weak social ties are as harmful to health as alcoholism~\cite{holt2010social}. For adolescents, the need for trusted interpersonal connections, especially with peers, is even more crucial than for adults~\cite{erikson_identity_1994, Davis2012-bq}. With 95\% of teens aged 13-17 actively engaging with friends on social media~\cite{noauthor_2023-xq}, the nature of these social interactions has fundamentally changed. As peer relationships increasingly occur online, it's essential to evaluate how technology supports teens' relationship-building.

Social Penetration Theory~\cite{altman1973social} explains that relationships develop through reciprocal and authentic self-disclosure, deepening over time~\cite{berger_social_2015}. However, social media often provides what has been termed ``empty calories of social interaction''~\cite{Prinstein2023-pd}. While these interactions may seem to meet adolescents' developmental needs, they often lack the substance necessary for genuine social benefits. Social media use can in fact exacerbate adolescents' developmental vulnerabilities, such as heightened sensitivity to social feedback and increased emotional reactivity due to ongoing brain maturation~\cite{Somerville2013-zf, Steinberg2005-qy, FalkEmilyB.Ph.D2014NRtE, PeakeShannonJ.2013Rase, SomervilleLeahH.2013TTBS}. This can hinder the formation of meaningful, supportive relationships. Despite these shortcomings, most teens regularly engage in extensive self-presentation on social media, sharing carefully curated content, an experience that can induce anxiety and distort users' perceptions of reality~\cite{Aacap_undated-ho}.

\textit{BeReal} is a social media platform designed to address the challenge of fostering genuine relationships online, where authentic interactions are often scarce. BeReal has a specific focus: ``\textit{everyday get a notification to share a genuine glimpse into your real life with the people you care most about}'' which they believe enables users to ``\textit{show your friends who you really are}''~\cite{noauthor_undated-pl}. Though BeReal was launched in 2019, long before the U.S. Surgeon General's 2023 advisory on social media and youth mental health~\cite{noauthor_2023-xq}, its design and mission align with the advisory's recommendation that tech companies ``seek to maximize the potential benefits and avoid design features that attempt to maximize time, attention, and engagement''~\cite{noauthor_2023-xq}.

On BeReal, users receive a notification informing them that they have a brief window of time to capture and share photos of their current activity and appearance, making strategic self-presentation nearly impossible. Consequently, the content shared on BeReal is more likely to be mundane or monotonous than content shared on other platforms. Yet, the app has become increasingly popular since early 2022, especially with youth~\cite{cycles_topic_nodate}. According to a 2023 Pew Research Report, BeReal usage among teens is comparable to that of Reddit or Twitch~\cite{anderson2023teens}. But regardless of the fate of the BeReal platform, it serves as a useful case study in designing to encourage authentic self-presentation. Self-presentation can build close relationships (when it is genuine) or provoke anxiety (when it is artificial). Thus, there is great value in creating spaces that encourage the former.

To the best of our knowledge, BeReal is the first major social media platform that explicitly purports to encourage authentic self-presentation. We sought to understand:
\begin{itemize}
    \item \textbf{RQ1:} How, if at all, does the design of BeReal influence users' self-presentation on the platform?
    \item \textbf{RQ2:} How do teens feel about self-presentation on BeReal?
\end{itemize}

To answer these questions, we conducted semi-structured interviews with 29 BeReal users aged 13 to 18, asking about their general experiences with BeReal. We found that BeReal's features and affordances are effective in encouraging users to share casual, uncurated snapshots of their daily lives. Features such as randomly timed daily notifications encourage authentic self-presentation via unstaged, unfiltered photo-sharing, and design elements like reciprocal posting encourage users to curate their audience, thereby allowing their self-presentation process to be less intentional or controlled. The participants saw many benefits of such designs. They especially appreciated that BeReal nudged users to collectively share more casual and less curated content, a change they said they sought on social media. However, some participants also felt that their ability to present themselves accurately and intentionally was too limited, and they explained that unfiltered photos taken at unplanned times are just one way of giving others insight into their lives. Others said that BeReal's explicit prioritization of ``authenticity'' in self-presentation sometimes led to toxic competition and judgment about which users are \textit{truly} being authentic.

In this paper, we provide empirical evidence of BeReal's success in encouraging adolescent users to present themselves authentically. This serves as a proof of concept that design can significantly influence self-presentation. We also document the limitations of BeReal's approach and outline design recommendations for creating a platform that encourages developmentally supportive online self-presentation for adolescents, such as scaffolding reciprocal sharing. Based on our findings, we introduce the Social Media Self-Presentation Matrix (SMSM). Grounded in theories on self-presentation goals and processes~\cite{Schlenker2003-yn}, this matrix presents ways that social media design can support different self-presentation strategies. It does so through four facets of self-presentation, spanning axes from Automatic to Controlled and from Self-Enhancing to Self-Expressive. Using the matrix, we highlight an untapped design opportunity to support Controlled and Self-Expressive self-presentation, which could offer an alternative way for social media platforms to achieve the goals initially intended by BeReal. Further, we propose a design agenda to support better relationship-building online based on authenticity for teens on social media. We hope this work will encourage designers and researchers to more comprehensively support authentic exchanges in teen peer-relationship-building on social media, addressing a developmental need that is currently unmet.
\section{Related Work}
\label{lab:rw}

\subsection{Adolescents and the Developmental Significance of Authentic Self-Disclosure}
Forming close, trusting relationships is fundamental to people's sense of meaning, happiness, and long-term well-being~\cite{argyle2013psychology}. For adolescents, this intrinsic need for trusted interpersonal connections, particularly with peers, is even more pronounced than it is for adults~\cite{Erikson1994-ld, Davis2012-bq}. As described by the Social Penetration Theory (SPT)~\cite{altman1973social}, interpersonal relationships develop from initial, surface-level interactions to deeper, more intimate connections through a gradual process of self-disclosure. The breadth and depth of these disclosures increase as the relationship strengthens and matures. Also central to this progression is the norm of reciprocity, which posits that mutual exchange is fundamental to these evolving relationships.

For self-disclosure to effectively enhance intimacy and trust in relationships, it must be authentic. Authenticity---the characteristic of being aligned with one's intrinsic cues rather than being influenced by external expectations or pressures~\cite{Peets2018-aq}---ensures that the information shared truly represents the individual's self, fostering genuine connections and understanding between people~\cite{Laurenceau1998-xi, Peets2018-aq}. Expressing their authentic selves and receiving acknowledgment and affirmation from peers is crucial for individuals, particularly adolescents, to develop a clear and coherent self-concept~\cite{app-generation}. The formation of a coherent identity, rooted in an authentic self that is recognized and validated by one's social circle, is fundamental to adolescent development~\cite{erikson_identity_1994}.
Research has shown that authenticity in self-disclosure is associated with life satisfaction~\cite{Bailey2020-od}, particularly among girls, linking authenticity to self-esteem and overall well-being~\cite{Impett2008-xm}. Furthermore, a higher degree of authenticity in self-disclosing communication is connected to enhanced self-perceptions, reduced feelings of isolation in school, and greater satisfaction in relationships among adolescents~\cite{Peets2018-aq}.

In recent years, particularly exacerbated by the COVID-19 pandemic and subsequent physical distancing measures, adolescents have increasingly turned to social media for social connection~\cite{Hamilton2022-xi, pitt2021kids}. As of 2022, nearly all teenagers aged 13-17 (95\%) own a smartphone or access the internet (97\%), with nearly half (46\%) using the internet ``almost constantly''~\cite{Vogels2022-jf}. Previous studies have identified that the primary motivation behind teens' use of digital media is to maintain connections with existing friends~\cite{ito2013hanging, Boyd2008-zi, Davis2012-bq}. Given the critical role of online platforms in adolescent lives and the value placed on authenticity in their relationships, it is understandable that modern teenagers actively seek exchanges of authentic self-disclosure on social media and broader media contexts~\cite{Darr2022-af, css_teens}. Further, the significantly reduced social cues on online platforms heighten uncertainty, making self-disclosure especially crucial for (adolescent) users to understand each other and strengthen their connections~\cite{Schlosser2020-jg, tidwell2002computer}. This underscores the importance of examining how technology can support or undermine the development and maintenance of these relationships, with a particular focus on the reciprocal exchange of authentic self-disclosure throughout the stages of social penetration.

\subsection{Authentic Self-Disclosure and Self-Presentation}
Given that self-disclosure, especially in online contexts, involves extensive interpersonal interactions, it can be likened to a performance~\cite{goffman2016presentation} or an exhibition~\cite{Hogan2010-oh} that symbolically communicates information about oneself to others~\cite{blumer1986symbolic}. As a result, self-presentation becomes an inherent part of this process. Self-\textit{disclosure} is defined as \textit{``verbally communicating personal information about the self to another person''}~\cite{forgas2011affective} (p. 449), and reflects communicating a factual representation of oneself, regardless of its effect on one's public image. In contrast, self-\textit{presentation} is defined as \textit{``the goal-directed activity of controlling information to influence the impressions formed by an audience about the self''}~\cite{schlenker2003carryover} (p. 871). Thus, self-disclosure typically aligns with authenticity, whereas self-presentation may suggest a lack of authenticity. However, these two are not mutually exclusive~\cite{johnson1981self}; revealing true aspects of oneself can still be compatible with self-presentation, as such revelations can also support a favorable public persona~\cite{Schlosser2020-jg}. \textit{We adopt the view in this paper that \textbf{self-presentation} is a strategic form of \textbf{self-disclosure} that predominates on social media, where users, acting as performers, anticipate positive outcomes from their disclosures.}

Although self-presentation is often perceived as strategic gamesmanship, this view does not capture its full complexity~\cite{Schlenker2003-yn}. In reality, self-presentation can be deeply ``authentic,'' as in, it typically involves a portrayal that is somewhat enhanced and idealized, yet genuinely believed by the person presenting it~\cite{brown2014self, greenwald1985whom, williams2008people}. Self-presentation often aims for accurate self-portrayal~\cite{baumeister_self-presentational_1982, cheek1983self, leary2019self, schlenker1980impression, schlenker2000strategic}. This approach may serve several purposes: to ensure others perceive one in a manner that elicits respect and treatment congruent with one's identity; to receive affirming feedback that addresses personal uncertainties; and to embody the principle that ``honesty is the best policy,'' fostering a sense of authenticity while mitigating risks associated with deception. Furthermore, intentional self-presentation does not inherently signify inauthenticity; Schlenker~\cite{Schlenker2003-yn} uses the analogy of how a skilled lecturer meticulously crafts their presentation to effectively convey genuine and significant information to their audience. This stands in stark contrast to a disorganized lecturer whose lack of preparation may come across as insincere or self-absorbed. Similarly, the careful arrangement of personal information may better reflect an authentic self-depiction as perceived by the individual)~\cite{schlenker2000strategic}. In other words, presenting accurate information requires as much skill in self-presentation as presenting inaccurate information if it is to achieve the intended effect on the audience.

Authentic self-presentation can manifest through either automatic or controlled cognitive processes. Automatic processes are those which 1) occur without conscious awareness, 2) require minimal cognitive effort, 3) do not require conscious monitoring, and 4) are initiated involuntarily without outside triggers~\cite{Schlenker2003-yn, bargh1996automaticity}. In such instances, self-presentation scripts may be triggered automatically by specific aspects of the audience or the context, with the individual often unaware of how deeply they are influenced by the social environment and their interpersonal objectives~\cite{jones1990interpersonal, schlenker1980impression, schlenker1985identity, schlenker2003carryover, tetlock1985impression}. Automatic self-presentation is most common in familiar environments among close acquaintances or friends, where individuals feel a sense of security and positive regard (e.g., while relaxing at home with closer friends)~\cite{tyler2012triggering, Schlenker2003-yn}. Self-presentation is also more likely to be automatic when it aligns with one's established self-image and personality traits~\cite{Schlenker2003-yn}. Conversely, self-presentation often involves controlled processes in situations where the stakes or the audience are significant or where the individual is unsure about the impression they wish to project~\cite{schlenker1996impression, schlenker2000strategic}. Under such conditions, individuals tend to deliberately focus on the impression they are making, meticulously planning and rehearsing their behavior~\cite{tice1995modesty, schlenker1996impression}. The caveat of having to deviate from their automatic mode of self-presentation is that individuals may need to exert more cognitive effort, which can detract from their performance or quality of life~\cite{vohs2005self, Schlenker2003-yn}.

\subsection{Challenges with Authentic Self-Presentation Online}
While authentic self-presentation is highly valued, particularly among teenagers, it can be particularly challenging to achieve on social media. One characteristic of computer-mediated communication (CMC) that makes it more conducive to self-presentation than face-to-face (F2F) interactions is its asynchronicity~\cite{Schlosser2020-jg}. In contrast to the often messy, unscripted, and spontaneous nature of F2F communication, social media allows for more controlled exchanges due to the ability to reflect, edit, and revise messages~\cite{turkle2016reclaiming}. Though self-presentation can also be spontaneous~\cite{schlenker2003carryover}, self-presentation is often a complex process, reflecting a dynamic interplay between the self and the audience within a specific social context~\cite{Schlenker2003-yn}. It involves not only expressing oneself but also responding to situational pressures and conforming to the identity expectations of others~\cite{schlenker1996impression}. It embodies aspects of an individual's self-concept, personality, social roles, and perceptions of audience preferences. Consequently, people often leverage the asynchronicity of CMC to craft their self-presentation online~\cite{vazire2004perceptions} carefully.

Online communication also differs from F2F interactions because it typically involves broadcasting messages to many people simultaneously rather than engaging in one-on-one exchanges~\cite{Schlosser2020-jg}. This broadcasting nature of social media interacts with adolescent characteristics in ways that significantly complicate self-presentation. Adolescents typically experience a heightened sense of self-consciousness relative to adults or younger children, leading to feelings of being under constant scrutiny, a phenomenon referred to as the \textit{imaginary audience}~\cite{Elkind1967-dw}. This feeling is amplified in online environments where audiences are typically larger and less transparent than in offline scenarios~\cite{Zheng2019-mq}. Moreover, adolescents' \textit{imagined audience} on social media, defined as ``\textit{the mental conceptualization of the people with whom [they] are communicating}``~\cite{Litt2012-ir}, encompasses individuals from diverse contexts such as family and school, mirroring the complex layers of audiences they engage with on these platforms. This sense of \textit{context collapse}~\cite{BoydDanahMichele2008Tooc} prompts adolescents to employ self-presentation strategies in online environments to manage the tensions that arise~\cite{Marwick2011-lr, Boyd2008-zi, Tao2023-an, Vitak2012-ne}. These strategies often involve sharing the \textit{``lowest common denominator''}~\cite{Hogan2010-oh}: content that is universally deemed appropriate by all groups within the imagined audience~\cite{Hogan2010-oh, Ullman1987-rj}. Consequently, relationship-building processes are also affected, as the breadth, depth, and reciprocity of authentic self-disclosure become restricted, and users disclose only superficial aspects of themselves~\cite{Hogan2010-oh} or disengage entirely from sharing~\cite{hargittai2008participation}.

The third distinguishing factor of many social media sites from F2F communication is the publicly visible and quantifiable nature of audience feedback, such as ``Likes'' and comments. The opportunity to present oneself on social media opens up new avenues for identity development, including experimenting with self-presentation and expression~\cite{WeinsteinEmily2017IDit}. However, online self-disclosure often involves risks and harms~\cite{Barta2021-yh, Schlosser2020-jg}, and adolescents, particularly those who are sensitive to rejection and self-presentation~\cite{Yau2019-ab}, justifiably feel more cautious about self-disclosure~\cite{ryan_toward_2019}. The heightened self-consciousness often drives adolescents to present themselves favorably online to maintain their social standing~\cite{Yau2019-ab, Haimson2021-qd, Hamilton2022-xi}. This leads to a tension between authenticity and self-presentation online known as the \textit{``online authenticity paradox''}~\cite{Haimson2021-qd}: people aim for authenticity in their personal life and online presentation, but doing so necessitates sharing negative experiences. Sharing these experiences requires vulnerability, which can either be risky or beneficial~\cite{Barta2021-yh}. Explicitly investing effort in communication~\cite{Kelly2017-sc} may be helpful, but interactions on current social media that revolve around ``Likes'' often fail to provide the necessary support to make users feel safe in those interactions~\cite{canady2023apa}.

Collectively, this line of work suggests a need for online spaces that support adolescent peer interactions in ways that are developmentally sensitive, i.e., providing support for authentic self-disclosure that leads to relationship development. Yet, research suggests that, currently, social media platforms instead prey upon the fears that adolescents experience, such as those stemming from their heightened sense of self-consciousness~\cite{bukowski1998company}. For example, photo sharing, a prevalent form of communication on social media platforms, can exacerbate self-presentation concerns~\cite{Yau2019-ab, zhao_2008}. The perception of attractiveness is often central to adolescents' self-concept and self-esteem, particularly among girls. These social media sites, with their visually oriented nature, the capability to carefully curate and manipulate visual images, and the existence of a peer audience that provides reinforcing feedback~\cite{Choukas-Bradley2020-zv}, often promote social comparison. This comparison is made with meticulously selected and edited photos of peers, celebrities, and ``influencers''~\cite{FardoulyJasmine2017Tioa}, leading to unhealthy social comparisons and an amplified emphasis on self-presentation. Furthermore, social media platforms frequently place a higher emphasis on metrics such as time on site and return visits than on the quality of relationships. This approach further encourages users to display carefully curated content for approval and partake in continuous social comparisons.

\subsection{Online Authenticity and Technology Design} 
While certain affordances~\cite{Evans2017-ji, Treem2013-ox, DeVito2017-by} of online communication---like asynchronicity, editability, visualness, and publicness---encourage manicured self-presentation, other affordances can promote authentic self-disclosure. Platforms and features such as Snapchat, Instagram Story, and BeReal support ephemeral sharing, which despite potential context loss~\cite{Cavalcanti2017-zl}, grants users greater control over their narrative and privacy~\cite{DeVito2017-by}. Ephemeral posts reduce self-consciousness, encourage casual and authentic exchanges~\cite{Bayer2016-fa, Choi2020-ku, kreling_feeling_2021, xu_automatic_2016, Chiu2021-pj}, and enable deeper interactions, playful interactions, and private engagements~\cite{phua_uses_2017, Choi2020-ku, Chiu2021-pj, xu_automatic_2016}. 

Besides ephemerality, the perceived degree of audience control and the (smaller) size of the network also support authenticity. Users often set aside online spaces for closer relationships, fostering authentic interactions within intimate friend circles~\cite{vaterlaus_snapchat_2016, piwek_what_2016, kreling_feeling_2021, Chiu2021-pj, Taber2020-kv, Huang2022-hl}. Finsta (i.e., ``Fake Instagram,'' or secondary Instagram accounts) users, for example, \textit{``reconfigure''}~\cite{Xiao2020-ce} Instagram to establish smaller, more intimate environments, enabling the sharing of unfiltered and emotionally expressive content~\cite{Xiao2020-ce}. While this active audience management can be demanding, it effectively promotes authenticity on social media~\cite{Xiao2020-ce}. Furthermore, pseudonymity, which can be found on platforms such as Tumblr, can foster a greater sense of trust among users and with the platform itself. This affordance has been shown to promote online authenticity, particularly benefiting marginalized individuals who seek safe spaces for genuine self-expression and disclosure~\cite{DeVito2018-vb, Davis2023-ft, Barta2021-yh, Schaadhardt2023-jj}.

Previous studies in Human-Computer Interaction (HCI) and Computer-Supported Cooperative Work (CSCW) have also sought to explore new ways of supporting relationship-building and authentic self-disclosure online. For instance, an analysis by Stepanova et al.~\cite{Stepanova2022-vh} investigated the design of 50 artifacts aimed at promoting \textit{``genuine''} feelings of connection in technologically driven systems. Their research identified nine design strategies that enable genuine connection, with affective self-disclosure being the most pertinent to our study. Numerous systems employing this approach have leveraged biofeedback sharing (e.g., signaling physiological activity such as heart rate and its associated emotion to a partner) in dyadic communication settings~\cite{Qin2021-fa, MentisHelenaM2014MSaY, LiuFannie2021SOUt, LiuFannie2019ASBo, GervaisRenaud2016TTOE, Frey2018BSBt}. Alternative strategies were catered to broader connections beyond peer relationships, such as linking with the unity of humanity~\cite{Howell2019-el}, engaging with strangers~\cite{Mitchell2017-dd}, or maintaining distant relationships~\cite{Robinson2020-im}. In addition, a substantial body of research exists that examines how technology can facilitate self-disclosure and relationship enhancement through the use of situational conversation prompts. For instance, Zheng et al.~\cite{Zheng2021-kr} developed PocketBot, a chatbot that nudges users to resume, enliven, and deepen conversations. However, research in this area has primarily focused on using conversational prompts to support relationships with strangers~\cite{Hu2022-zx}.

Finally, several studies have explored BeReal from authenticity and/or relationship angles~\cite{Vanhoffelen2023-kx, Tirocchi2023-zp, Bulchand-Gidumal2024-lz, Snyder2024-jr, Maddox2023-ma}. First, Tirocchi~\cite{Tirocchi2023-zp} discusses how BeReal's authenticity aligns with the values of users aged between 20 and 23, showing that young people's media choices, including BeReal, reflect a desire for authenticity. Maddox~\cite{Maddox2023-ma} highlights how BeReal diverges from performance-based authenticity by emphasizing the promotion of \textit{``authenticity-as-realness''} through spontaneity. The authors note how the app's \textit{``panopticism''} and self-monitoring are what enable users to put realness before performance, a culture not seen previously on other social media platforms. Taylor~\cite{Taylor2023-rz} analyzed BeReal's designs, finding that they encourage users to engage with \textit{``sporadic''} authenticity. On the other hand, Snyder~\cite{Snyder2024-jr} critiqued BeReal's claim to authenticity by examining its user experience and rules that enforce an \textit{``always-on mentality,''} arguing that this approach increases the need for external curation instead of promoting true self-expression. Finally, Reddy and Kumar~\cite{Reddy2024-ic} explored how BeReal's design features shaped what constitutes ``authenticity,'' finding that BeReal's affordances align more with \textit{``extemporaneous interaction''} than \textit{``comprehensive self-presentation.''} Despite extensive research on BeReal and online authenticity, there remains a limited understanding of how BeReal's design aligns with adolescent developmental needs, particularly regarding self-presentation strategies and relationship-building on social media. Our study fills this gap by analyzing BeReal's approach beyond affordances, examining how its design cultivates specific cultural norms and expectations of authenticity that resonate with teens' heightened sensitivity to peer perceptions. We also explore how these factors contribute to both engagement and potential toxicity in self-presentation practices.
\section{Background}

\begin{figure}[htbp]
    \centering

    \begin{subfigure}[b]{\linewidth}
        \includegraphics[width=\linewidth]{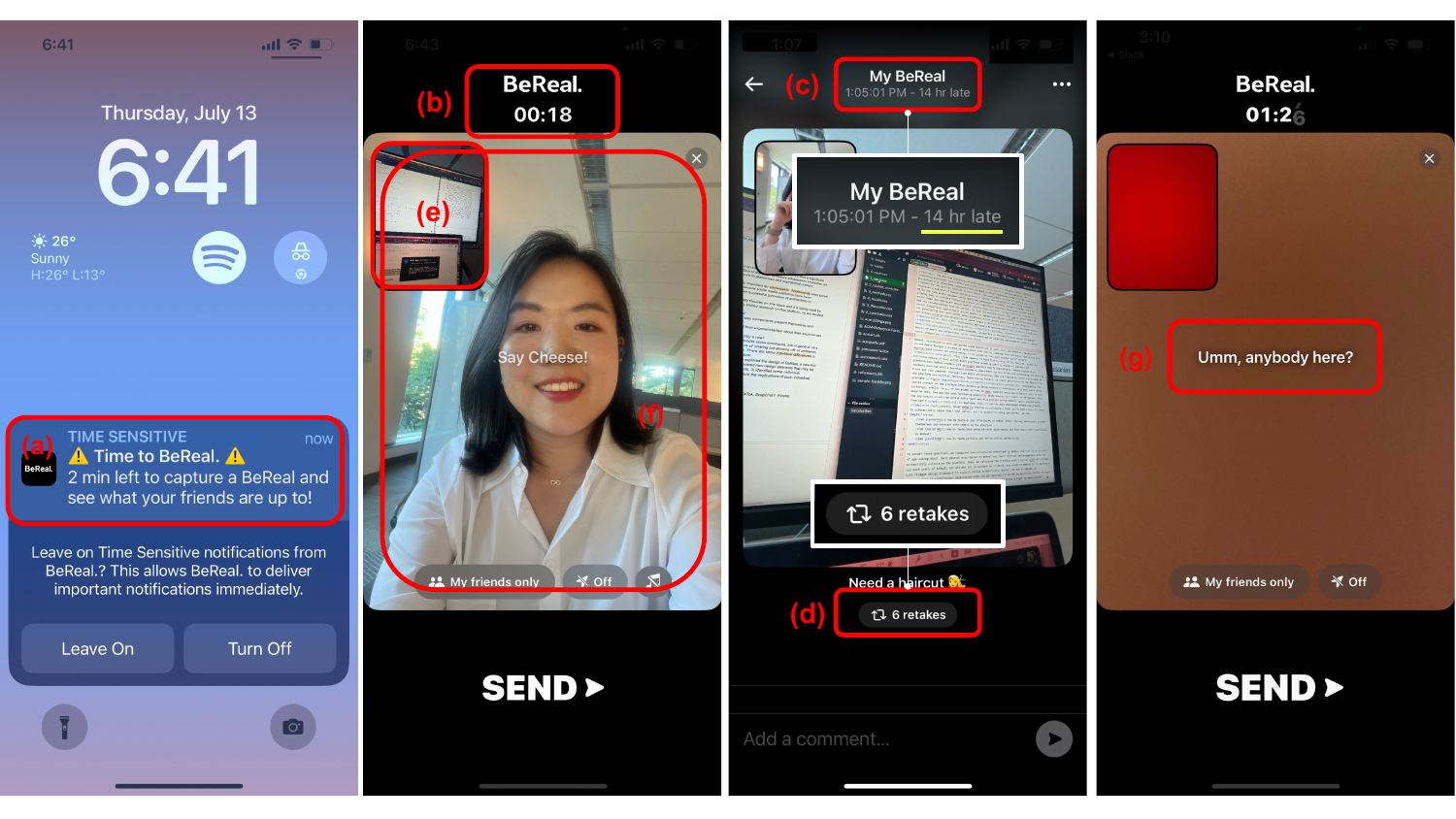}
    \end{subfigure}
    
    \begin{subfigure}[b]{\linewidth}
        \includegraphics[width=\linewidth]{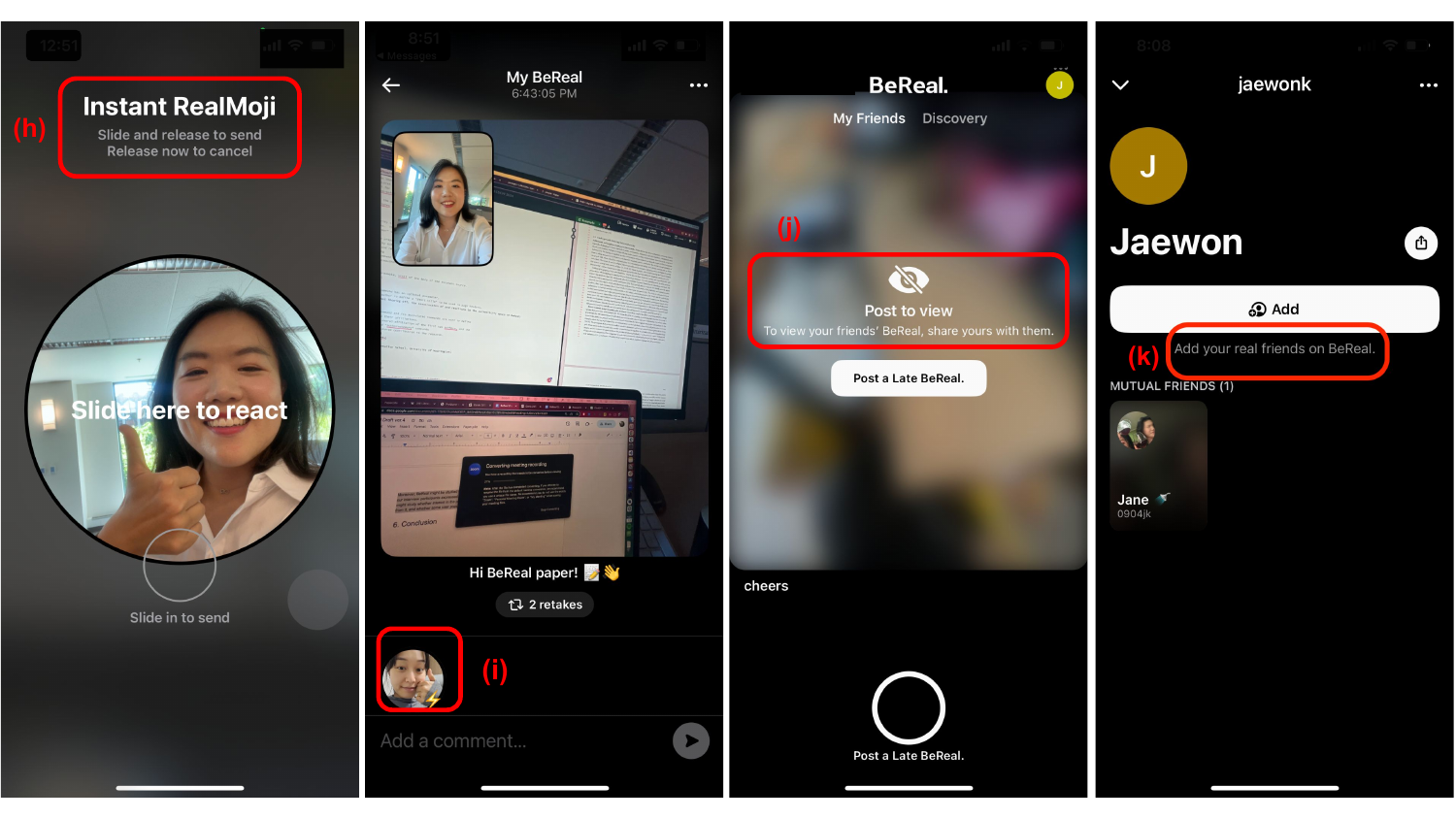}
    \end{subfigure}
    
    \caption{Screenshots of BeReal features: (a) randomly timed daily notification; (b) a 2-minute countdown timer for BeReals taken on time; (c) timestamp for a late BeReal with a ``late'' status; (d) the number of retakes; (e) \& (f) front and back photos; (g) feedback on photo encouraging users to show their faces; (h) Instant Realmoji that requires users to take a photo each time; (i) Instant Realmoji used on a post; (j) text explaining reciprocal posting; (k) text explicitly encouraging users to add closer friends}
    \Description{Eight screenshots of BeReal, four at the top row and four at the bottom. Going in the order from the top left to the bottom right: The first image (a) shows a phone screen at 6:41 AM on Thursday, July 13, with a ``Time to BeReal'' notification at the bottom. The notification indicates that the user has 2 minutes left to capture a BeReal, with options to leave notifications on or turn them off. In the second image (b), the BeReal app is open, showing a live selfie of a woman smiling in an indoor space. A small inset image (e) in the top left corner captures the rear camera’s view of a computer screen. A timer at the top (b) reads ``00:18.'' The selfie (f) occupies most of the screen, with privacy options (``My friends only'' and ``Off'') and a ``SEND'' button visible at the bottom. The third image (c) shows a BeReal posted 14 hours late, with a timestamp of 1:05:01 PM. The BeReal includes an image of a computer screen with text, and the bottom of the image displays ``6 retakes'' (d) with the same privacy and sharing options as the previous image. In the fourth image (g), the BeReal shows a blank or blurry image with a red square in the upper left. A message in the center reads ``Umm, anybody here?'' The privacy settings and ``SEND'' button are at the bottom, similar to the other images. In the fifth image (h), the screen shows an ``Instant RealMoji'' feature with the prompt, ``Slide and release to send.'' A large circular preview displays a selfie of a woman giving a thumbs-up gesture. The instruction ``Slide here to react'' is written over the selfie, with a slider at the bottom prompting the user to send the reaction. The sixth image shows the user's BeReal post, timestamped at 6:43:05 PM. The BeReal captures two images: a selfie of the user in the upper left corner and a view of their computer screen in the main frame, showing a document being worked on. At the bottom, a comment reads, ``Hi BeReal paper! (paper emoji)(wave emoji),'' with a ``2 retakes'' indicator visible. A small circular image of the user’s face (i) is positioned in the bottom left, next to the comment section. In the seventh image (j), the screen displays a blurred background with a message in the center: ``Post to view. To view your friends' BeReal, share yours with them.'' Below this is a ``Post a Late BeReal'' button, prompting the user to submit their BeReal to unlock their friends' posts. In the eight image (k), the user's BeReal profile is shown. The profile displays a yellow icon with the letter ``J'' for the username ``Jaewon.'' Below the username, an ``Add'' button is visible with the instruction, ``Add your real friends on BeReal.''}
    \label{fig:screenshots}
\end{figure}

\begin{table*}[!h]
\caption{BeReal Features}
\Description{Table with descriptions of BeReal's key features as of December, 2022.}
\label{tab:features}
\resizebox{\textwidth}{!}{%
\small
\begin{tabularx}{\textwidth}{p{0.17\linewidth} X p{0.12\linewidth}}
\toprule
\textbf{Design} & \textbf{Description} & \textbf{Image} \\
\midrule
Daily notifications at random times & Once at any moment of each day, the app sends out a notification with the message, ``Time to BeReal. 2 min left to capture a BeReal and see what your friends are up to!'' & Figure \ref{fig:screenshots}a  \\
\hline
Two-minute posting window & Clicking on the daily notification initiates a two-minute countdown or an upward counting of how late the post is, shown above the camera & Figure \ref{fig:screenshots}b \& c \\
\hline
Retake counter & Users can view the number of times any photo, including a friend's photo, was retaken before it was posted & Figure \ref{fig:screenshots}d \\
\hline
Ephemerality & BeReal posts, comments, and reactions remain visible to other users until the subsequent notification is sent on the following day, indicating that another post is due, while the posters themselves retain continuous access. & \\
\hline
Front and back photoshoot & All BeReal posts require that users capture images from both the front and rear camera almost at the same time & Figure \ref{fig:screenshots}e \& f \\
\hline
No filter & The in-the-moment pictures taken on BeReal cannot be modified, and no filters are available. Users are not able to choose past images from photo albums either. & \\
\hline
Photo feedback & After the first photo is taken with either the front or back camera, users wait for the second photo to be taken with the opposite camera. During this interval, the platform provides feedback on the initial image. The feedback exerts subtle pressure encouraging users to show their face and smile, for example, ``Don't move and smiiiile,'' and ``Wait, wait, wait, now smile.'' & Figure \ref{fig:screenshots}g \\
\hline
Realmojis & Realmojis enable users to respond to posts with a photo of their own instead of a standard ``like'' or ``heart.'' There are six options: thumbs up, happy face, surprised face, heart eyes, crying while laughing, and a lightning bolt. The first five, or ``pre-made'' (P18) Realmojis, utilize stored photos that users have taken at some point to represent each of these five emotions. The lightning bolt is an ``Instant Realmoji'' that requires a new photo each time. & Figure \ref{fig:screenshots}h \& i \\
\hline
Reciprocal posting & Users on BeReal can only view other people’s posts if they have contributed their content for the day & Figure \ref{fig:screenshots}j \\
\hline
Chronological feed & Unlike most mainstream social media platforms, BeReal’s feeds are strictly chronological. Consequently, users see posts from all their BeReal friends, regardless of how frequently they interact or their relationship strength. & \\
\hline
Friend count hidden & BeReal does not highlight the number of friends a user has. Users cannot see the friend count of others. Concerning their own friend count, they will see how many friends they have until they surpass 50 friends, at which point their friend count will be displayed as ``50+.'' & \\
\hline
Friction for adding new friends & The platform explicitly promotes the intentional addition of friends with statements such as ``On BeReal connect with your close friends'' or ``Add your real friends on BeReal'' beneath the friend request button. Unlike other platforms, which make numerous friend suggestions, BeReal only syncs local phone contacts and does not allow user searches. & Figure \ref{fig:screenshots}k\\
\bottomrule
\end{tabularx}
}
\end{table*}

BeReal incorporates several design elements that funnel users towards creating a space for sharing unfiltered photos with a closer circle of friends. For instance, BeReal users receive daily notifications that encourage frequent, low-stakes posting. These notifications are sent at random times each day, as determined by the system, prompting users to post photos from both their front and back cameras within a two-minute window. Although users can postpone their response to the notification and post at a later time, such posts are tagged as ``late.'' Notifications about these ``late'' posts are sent to other users, and the delay in posting is highlighted above the post, e.g., ``14 hr late.'' The number of retakes is also visible to friends. While taking the photo, users are nudged to show their faces with prompts like ``Your friends will definitely prefer to see your face!'' and ``Umm, anybody here?'' Users can react to posts with ``Realmojis,'' emojis created from photos of their own facial expressions. These two-photo posts are called ``BeReals,'' and users are unable to view others' posts until they have shared their own BeReal, and posts in feeds are displayed in chronological order. BeReal does not feature like counts or friend/follower counts. While users can see their own number of friends, any number exceeding 50 is simply displayed as ``50+.'' The platform also explicitly promotes adding ``real friends'' on BeReal. A detailed overview of these features, as of December 2022, when the study interviews were conducted, is presented in Table \ref{fig:screenshots}.

\section{Method}
We conducted semi-structured interviews with 29 current BeReal users. All participants had used BeReal for at least one week, had at least one friend on the platform, and had posted at least three posts. Interviews were conducted over a period of two months, with 14 interviews taking place at the end of December 2022 and an additional 15 at the end of January 2023. The procedure was approved by our Institutional Review Board (IRB).

\subsection{Participants and Recruitment} We used a combination of convenience and snowball sampling to recruit participants. Initial recruitment efforts involved reaching out to any known BeReal users aged between 13 and 18, as well as contacting the university's research participation pool to seek potential participants. Three participants were recruited from this effort. Participants were offered an additional \$10 incentive for recruiting friends who met the study's inclusion criteria. Three additional participants were recruited this way. To reach a broader audience, we posted ads on Instagram and Facebook targeting 13-18-year-olds residing in the United States. A total of 195 individuals responded and completed a screener survey. From this pool, 46 respondents met our inclusion criteria. We scheduled and conducted interviews with 29 of these respondents on a first-come, first-served basis.

The mean age of the participants was 15.6 years (SD=1.54), with 22 (75.9\%) identifying as girls, 6 (20.7\%) as boys, and one (3.4\%) as non-binary. The participants' self-described racial/ethnic breakdown included 11 (37.9\%) White/Caucasian, 9 (31\%) Asian, 3 (10.3\%) African American, 2 (6.9\%) Hispanic, 1 (3.4\%) Asian/White, 1 (3.4\%) Hispanic/White, and 1 (3.4\%) Asian/African. On average, participants reported having used BeReal for 5 months (mean=4.8 months, sd=1.7 months) at the time of the interview, excluding two novice users who had used BeReal for less than one month. To denote novice users with less than one month of experience with BeReal, add an ``n'' suffix to their participant name (e.g., ``P10'' as ``P10n''). Approximately 58\% of BeReal users in the U.S. identify as women~\cite{Curry2022-ha}. We do not have the racial or ethnic data for the overall population of BeReal users. The full list of participant information is available in Table \ref{tab:participants}.
\begin{table*}[!h]
\caption{Participant Demographics}
\Description{Table with information on the participant ID, age, gender, ethnicity, time since BeReal install, and number of friends on BeReal}
\label{tab:participants}
\begin{tabularx}{\textwidth}{@{}l l l X X X@{}}
\toprule
\textbf{PID} & \textbf{Age} & \textbf{Gender} & \textbf{Ethnicity} & \textbf{Time since BeReal install} & \textbf{No. of friends on BeReal} \\
\midrule
    P01 & 14 & Boy & White/Caucasian & 3 months & 30 \\
    P02 & 13 & Girl & White/Caucasian & 2 months & 22 \\
    P03 & 14 & Girl & White/Caucasian & 3 months & 20 \\
    P04 & 17 & Girl & White/Caucasian & 5 months & 25-30 \\
    P05 & 13 & Girl & African American & 2 months & 20-25 \\
    P06 & 16 & Girl & White/Caucasian & 3 months & 50+ \\
    P07 & 17 & Girl & White/Caucasian & 4 months & 50+ (150)** \\
    P08n & 17 & Non-binary & White/Caucasian & 1 week & 1 \\
    P09 & 16 & Girl & Asian/Asian American & 5 months & 50+ \\
    P10n & 18 & Girl & African American & 1 week & 8 \\
    P11 & 17 & Boy & White/Caucasian & 6 months & 50+ (70-80) \\
    P12 & 17 & Girl & African American & 4 months & 50+ (70) \\
    P13 & 14 & Boy & White/Caucasian & 4 months & 20-30 \\
    P14 & 16 & Girl & Asian/Asian American & 4 months & 20 \\
    P15 & 15 & Girl & Asian/Asian American & 4 months & 30 \\
    P16 & 13 & Girl & Asian/White & 5 months & 3 \\
    P17 & 15 & Girl & White/Caucasian & 6 months & 30 \\
    P18 & 15 & Girl & Asian/Asian American & 6 months & 15 \\
    P19 & 17 & Boy & Hispanic & 7 months & 20-40 \\
    P20 & 18 & Boy & White/Caucasian & 7 months & 50 \\
    P21 & 17 & Girl & Asian/Asian American & 5 months & 30-50 \\
    P22 & 15 & Boy & Hispanic & 4 months & 10-15 \\
    P23 & 16 & Girl & Asian/Asian American & 6 months & 50+ \\
    P24 & 16 & Girl & Hispanic/White & 8 months & 50 \\
    P25 & 17 & Girl & Asian/Asian American & 8 months & 70 \\
    P26 & 15 & Girl & Hispanic & 5 months & 50+ (100+) \\
    P27 & 13 & Girl & Asian/Asian American & 6 months & 50+ \\
    P28 & 15 & Girl & Asian/African & 5 months & 15 \\
    P29 & 17 & Girl & Asian/Asian American & 2 months & 10 \\
\bottomrule
\end{tabularx}
\begin{flushleft}
\small
\textit{Notes:} Novice users are denoted with (n). Numbers in brackets (N) for 50+ friends is the participants' estimate of the number of friends they have.
\end{flushleft}
\end{table*}

\subsection{Procedure} One member of the research team conducted 29 semi-structured Zoom interviews. The interviews lasted between 30 and 45 minutes. Participants received an Amazon gift card worth US\$20 as a thank-you for their time. The interview prompts probed 1) participants' general experiences with using BeReal, 2) their engagement with other social media platforms, 3) their motivations for using BeReal, 4) how they engage with its features, and 5) their perceptions of online authenticity (see Appendix \ref{appendix:protocol} for the complete interview protocol).

\subsection{Data Analysis} We conducted a reflexive thematic analysis~\cite{braun2019reflecting} of the 29 interview transcripts. We chose this method of analysis as it is a structured yet flexible approach that is effective at generating a nuanced view of the data, aligns with a constructionist paradigm, and is not tied to a pre-existing theoretical framework, which is appropriate for this exploratory work. We began the analysis with open coding in three phases over the course of 2.5 weeks. In the first phase, the first author performed line-by-line coding of the transcripts. Codes during this phase were descriptive and close to the data, for instance, ``weaponizing authenticity,'' ``tech empowerment,'' and ``casual, low-pressure.'' In the second phase, all five members of the research team individually coded the same transcripts line-by-line using the Delve Tool~\cite{delve}. The aim of this phase was to scrutinize the data as closely and comprehensively as possible. The third phase involved synthesizing the codes from the first two phases to construct higher-level themes. The five authors used the Miro Board~\cite{miro} tool for affinity diagramming to visually group related codes into potential themes. The first author then synthesized these potential themes into a single set of proposed themes. Examples of such themes included ``features and affordances that encourage or discourage inauthenticity,'' ``cultures and norms of authenticity on BeReal,'' and ``adolescents' interactions with the features and culture of authenticity.''

Over the next two months, the research team met every 1-2 weeks to discuss, refine, and iterate on the codes and the constructed themes. At the end of these iterations, we had a finalized codebook with 43 refined codes, which were grouped into four overarching themes. After finalizing the codebook, four authors, excluding the first author, divided the 29 transcripts among themselves and coded different transcripts. The first author coded all of the transcripts. This process ensured that at least two members of the research team reviewed each transcript. Once all transcripts were coded, the authors who coded the same transcripts met to discuss and resolve any coding disagreements.

\subsection{Ethical Considerations and Researchers' Positionality}
As academic researchers focusing on teen social media behaviors, particularly their use of the BeReal app, we recognize our positionalities as adults with limited firsthand experience of the current social media landscape as it is lived and navigated by teenagers. No one on the research team has been an active user of BeReal, and our engagement is limited to only two friends on the platform. This distance from the research subject(s), both in terms of generational gap and practical experience, fundamentally shapes our approach to this research.

From the outset, we positioned ourselves not as experts but as learners, genuinely curious about the nuanced ways in which teenagers engage with social media platforms like BeReal. We approached our participants with openness, explicitly communicating our lack of affiliation with BeReal and emphasizing that honest, authentic responses were invaluable to our understanding. By framing ourselves as outsiders looking to be educated, we hoped to mitigate any power dynamics inherent in the researcher-participant relationship and encourage a more forthright sharing of experiences.

The ethical considerations in conducting research with minors on sensitive topics such as social media behavior were paramount throughout our study. We obtained both written and verbal consent from all participants\footnote{Consent from their parents was waived due to the practicality of recruiting.}, providing options for anonymity and comfort, such as turning off cameras during interviews. Our communications, especially when seeking clarifications via email, were crafted with care to be clear, respectful, and mindful of the participants' privacy and well-being.

The perspectives that participants shared with us suggest that our approach succeeded in yielding rich, authentic perspectives. Teens shared critical reflections on the norms of BeReal, admissions of judgment towards others, expressions of self-consciousness, and deviations from expected engagement patterns.

\section{Results}

\subsection{How BeReal's Design Encourages Authentic Self-Presentation}
\label{section:enforce}

\begin{displayquote}
\textit{``I feel like not showing your face \ldots that's one of the parts of what BeReal's used for. They use their front-facing camera and the back-facing camera\ldots [and have] little prompts that pop up saying `smile' and whatnot. It just implies that you should show your face, and then next photo is showing what you're doing right now\ldots Same thing with the reaction emojis, that you react with your face, and if you're just not showing your face on the platform, it just defeats the purpose of that front and back camera being used and the reaction emojis being used.''} - P22
\end{displayquote}

For the most part, participants did \textit{not} join BeReal because they were seeking a platform that promotes authentic self-presentation. Instead, participants said they joined for social reasons, saying things like, \textit{``all my friends were on it''} (P03) or \textit{``my friend told me to''} (P09). Others were intrigued by the platform after coming across it on Instagram or TikTok. However, participants shared that once they became BeReal users, the platform's features influenced their self-presentation and led them to share frequently and ``authentically'' (specifically, to post unfiltered, unstaged photos of themselves). Participants identified four key ways that BeReal's design promotes authentic self-presentation: 1) discouraging staged posts, 2) relieving posting pressure, 3) encouraging interactions with smaller social circles, and 4) maintaining a simple, mission-focused app design. We describe each of these in more detail below. 

\subsubsection{Designs that discourage staging.} BeReal has many features not present on other social media platforms that explicitly discourage users from staging their photos. First, the random timing of notifications deters staged posts. As P22 said, \textit{``I think it's just the whole concept of the app, BeReal---it randomly going off. It's supposed to catch you off guard.''} They explained that the two-minute posting window encourages spontaneous posts, saying things like, \textit{``you don't have time to get ready or to make yourself look good \ldots nobody can stage something in two minutes''} (P25). Similarly, P03 commented, \textit{``You can't really fake it because it's like a two-minute thing.''} Separately, participants explained that the retake counter further discourages staging by creating subtle pressure to minimize retakes. They said things like, \textit{``People didn't want other people to think that they were retaking all their photos and stuff''} (P26). If a face is not detected in the frame, the platform will nudge the user to show themselves, with messages like \textit{``Your friends will definitely prefer to see your face!''} and \textit{``Umm, anybody here?''} As P13 shared, \textit{``When you take the photo and it just puts, like, the text up on the screen like, like `great smile' or whatever, or like if you take an all-black photo it says like, `where are you?' so \ldots it gives you a good impression of [what BeReal is about] just based on those kind of little details.''}

Participants explained that these features funnel them into very specific self-presentation choices. Some participants shared that they retake photos only when their \textit{``eyes are closed''} or they \textit{``accidentally put [their] finger in the camera''} (P11), not due to self-consciousness. They also said they avoid posting blank images, aiming to \textit{``always [have] something there''} (P26), because blank posts \textit{``defeat the whole point''} (P20). They said they try to post as soon as they receive a notification because waiting to post renders it \textit{``kind of pointless''} (P04). They described the way BeReal encourages reciprocity and mutual sharing, saying, \textit{``You can't see anyone else's BeReal until you post yours for the day,''} likening it to an \textit{``exchange''} (P04). Given this context, posting a blank photo---withholding a piece of their lives---while being able to view others' posts seemed like\textit{``cheating''} (P13), \textit{``hypocritical''} (P13), and \textit{``not fair''} (P17) to some participants. Further, posting blank photos was seen as undermining or defeating the \textit{``point''} (P15) or \textit{``purpose''} of the app (P22). Participants said that when they see others post blank photos, their reactions range from mild disappointment to more pronounced disapproval. Collectively, participants made clear that BeReal's features communicate specific expectations for how they should engage with the platform. These designed expectations lead them to post unstaged photos and expect other users to do the same.

\subsubsection{Designs that take pressure off posting.} Unlike most mainstream social media platforms, where sharing edited and curated content is the norm, BeReal's designs allow users to share photos without the pressure to curate their posts by making sharing more low-stakes. For example, participants described the app-defined cadence of the daily posting by saying, \textit{``I think it does feel different just because I'm doing it so frequently, and because it has this connotation of, `okay, this doesn't matter'''} (P24). Ephemerality further lowers the stakes for engaging with others. P16 shared: \textit{``I don't feel as pressured to make it look super nice because it's like it'll just be away in 24 hours.''} Moreover, the effort Realmojis demand and their ephemeral nature reduce the pressure to collect them. Unlike other platforms where ``Likes'' signify social status, on BeReal, they do not carry much weight. As P15 put it, \textit{``There's like no `Likes'\ldots I don't think people usually really compare [the number of Realmojis they received].''}

The effect of these features is that even compared to other casual online spaces like Snapchat, Instagram Close Friend Stories, and Finsta~\cite{Xiao2020-ce}, participants found BeReal to be less crafted, unlike other platforms where users put more effort into curating their content. P18 shared that they post photos on BeReal that they would not share on their Finsta because they are \textit{``just too boring.''} P04 stated that Close Friends Stories were also considered more curated than BeReal posts, as they are primarily used for sharing funny or noteworthy personal events. Because all users share whenever \textit{``the app tells [them] to do so''} (P25), users feel comfortable posting the \textit{``least exciting parts''} (P27) of their day. P23 clarified the effect of the frequent, low-fidelity posting on BeReal stating, \textit{``[on other social media platforms] you have a purpose in posting these [posts]. BeReal, you're not supposed to have a purpose in posting. You're just posting just because of a random notification, throughout the whole day, would just tell you to post. There isn't that intent behind it.''} One participant used the analogy that BeReal \textit{``feels a lot like a Polaroid,''} whereas Instagram \textit{``would be a picture hung on your walls---perfect and made to show off''} (P29).

\subsubsection{Designs that encourage audience curation.} Context collapse inevitably leads to curated self-presentation, and BeReal attempted to address this phenomenon by encouraging users to curate their \textit{audience} instead of their posts. Specifically, BeReal's features encourage engagement with closer acquaintances and friends. This approach makes it easier for users to post casual, daily content without the fear of judgment, fostering an environment that promotes casual sharing rather than curated self-presentation. By enforcing reciprocity, the platform encourages users to limit their BeReal circles to those with whom they feel comfortable sharing their daily lives. As a result, users typically have fewer friends on BeReal than they do on other platforms. P04 explained that they add only the people they \textit{``talk to regularly or that someone that I would want to stay in touch with''} because they wish to feel \textit{``comfortable with [friends] seeing parts of [their lives] that are personal.''} Also, according to P10, scrolling through posts from less familiar friends on BeReal can become \textit{``overwhelming and just sort of boring.''} Therefore, users tend to curate their friends, adding only those \textit{``that they care about seeing what they're doing every single day''} (P04). Moreover, by hiding friend counts, BeReal discourages popularity contests that can hinder authentic sharing. P04 explained: \textit{``On BeReal there's definitely the least amount of pressure because nobody can see how many friends someone else has. Like, I don't even know how many friends I have exactly\ldots the number really doesn't matter so you can have as few or as many as you want.''} Overall, participants expressed that their friend lists on BeReal, in general, are \textit{``pretty much invite-only''} (P22).

These features, which encourage audience curation, result in BeReal being described as \textit{``friendly''} (P26), \textit{``cozy''}, \textit{``home-like''} (P16), and \textit{``communal''} (P04). P18 noted, \textit{``very few people will see my BeReal\ldots it feels like there's less pressure.''} Participants also said that the BeReal experience made them \textit{``feel connected''} (P28) to other users and appreciated the platform's focus on updates from close friends that felt like \textit{``check-ins''} (P14). Participants described BeReal as a \textit{``judgment-free''} (P16) environment that feels like a \textit{``safe, close space''} (P29) fostered by a carefully curated audience as a defining characteristic that sets BeReal apart from other social media platforms.

\subsubsection{Designs that focus on a single purpose.} As of December 2022, BeReal maintained a minimalist interface design, featuring only elements that serve its single purpose of sharing candid photos. This minimalism clarifies BeReal's mission of authentic sharing with the users. For instance, participants viewed the act of retaking a photo as somewhat discordant with the platform's mission and explained: \textit{``the point of BeReal is like you take one photo and you're done''} (P07).

P15 emphasized the platform's focused mission, suggesting that expanding BeReal's feature set could dilute its essence: \textit{``For BeReal, I feel like it's the type of social media that's a bit more hard to expand because expanding it, I feel like they would have to give up the whole point of BeReal. It's hard to have both, where it's a very authentic app, and have people creating content because they want others to view it.''} Adding more features to BeReal would make it less distinct from other social media platforms. For example, P09 shared that \textit{``if you were to take multiple pictures, it would kind of become Instagram.''} P23 echoed this sentiment, saying that \textit{``Instagram is complicated because there's edits''} and that BeReal is \textit{``supposed to be plain, simple, minimalistic.''} P22 felt that Instagram's additional features create \textit{``social hierarchy''} but explained that on BeReal \textit{``there's not a whole lot \ldots to be judgmental of.''}

The simplicity of BeReal not only makes its purpose clear to users but also sets the platform apart from features like TikTok Now~\cite{TikTok2022-fe}, which, despite its striking resemblance to BeReal, is perceived differently because it is integrated into TikTok's broader functionality. For example, P25 expressed that \textit{``I feel like most people know TikToks as those short 15-second videos, whereas BeReal's its own thing. And so to merge [Now into TikTok] just feels a little bit more awkward and then more clunky, and then it just doesn't feel as real as BeReal.''} Also, P24 shared that the audience in each app is different; \textit{``TikTok has more functions, and I guess that you can't really curate the friends that you interact with on TikTok as much as you can with BeReal.''} TikTok Now is a part of TikTok, which to the participants has \textit{``always been a general video sharing app''} and \textit{``it's the same concept as BeReal but \ldots they don't talk as much about like being real, that's not really their focus the same way that it is with the BeReal app''} (P08n).

\subsection{The Perceived Benefits of BeReal's Design} 
\label{benefits}

\subsubsection{Validating an unfiltered life} \label{bereal_authentic} Most participants expressed appreciation for BeReal's mission to promote more authentic self-presentation. For example, P12 stated that it is \textit{``hard to like, eliminate that kind of idea of everyone and wanting to always appear like perfect and put together, but I think BeReal is doing a good job at trying to like begin and combat adding that type of thing.''} P02 reflected on the impact of the authenticity norm on BeReal, noting a shift in how they view social comparisons: \textit{``There's a lot of girls in my grade who I follow. And, like, at school, they present themselves as these confident \ldots always feeling fabulous \ldots and then seeing them in bed eating snacks, watching Netflix, and you're like they're a normal human being \ldots they aren't perfect all the time.''} This brings out a sense of \textit{``normalcy''} to unfiltered sharing as P28 succinctly put it, and offers a view of \textit{``their daily life rather than their glamorized events''} (P25).

The contrast between the parts of themselves that people share on BeReal versus Instagram was particularly striking to participants. For example, P24 observed that BeReal is \textit{``breaking down the walls that Instagram has put up since it was launched. You're seeing people at their best, and you're seeing people edited. BeReal is trying to get people to understand that that's not real.''} This aligns with P06's assessment, who shared that \textit{``all I see of them on Instagram is like traveling''} and that they find it \textit{``refreshing''} that on BeReal, they can see their friends \textit{``staying at home''} (P06). P24 further explained: \textit{``If Instagram is a room of people at their best moment with heavy editing, BeReal is trying to get people at their worst moment with no chance of editing. BeReal is anti-Instagram, and Instagram is just people trying to show off their best selves.''}

Participants viewed BeReal's emphasis on an unfiltered life as indicative of a broader potential shift in social media culture. For example, P25 described BeReal as \textit{``trying to bridge that gap between social media and real-life''} by \textit{``bring[ing] more intimacy into the social media world,''} and they enjoyed being a part of \textit{``a collective effort to like destigmatize like having unfiltered moments.''} Further, P08 expressed the platform's mission as \textit{``hopeful and inspiring \ldots it was created to help teach people that like you don't need to stress all the time about social media and being cool and like the way that people perceive you. It's okay to be your real self.''} P24 expressed appreciation for BeReal's new take on social media by acknowledging how BeReal reduces the pressure, workload, and stress that come with posting on other platforms, stating \textit{``There's so many people that overthink it and give just an Instagram post too much value and platforms like BeReal are it doesn't matter. It's just an app \ldots That lightheartedness is really valuable in terms of social media.''}

\subsubsection{Motivating positive changes} A few participants also said that using BeReal encouraged them to be the person they aspire to portray online. For example, P02 stated: \textit{``I'll feel guilty if I am not, like, not getting out of bed or if like, my BeReal notification comes along and I'm still in bed or something, but most of the time it's, like, it's nice like when I'm doing something productive, and then my BeReal comes along, and I'm like, `oh cool I get to be doing something interesting while I take my BeReal.'''} Rather than feeling overwhelmed by this, P02 stated that they feel \textit{``motivated''} and that they want to be productive when their BeReal comes along. P24 echoed this perspective by saying that looking back through their BeReal posts served as \textit{``a little mirror''} to them.

Moreover, many expressed that, despite initial self-consciousness over seemingly mundane or unflattering posts, BeReal has empowered them to embrace and share their genuine selves. For example, P11 shared that they \textit{``kind of like to lean into the discomfort because it feels more honest.''} P18 shared similar sentiments about overcoming self-consciousness, stating, \textit{``Well, I think about how it is important to be realistic on social media, and I think about how most people who see it wouldn't really care that much \ldots, so I try to just ignore it and press `post' and be casual about it.''} P02 echoed this sentiment, saying that \textit{``BeReal allows you to be authentic and kind of share what you're going through. In an open way and like you're supposed to be vulnerable, maybe, or you're supposed to be your true self. So I think BeReal in some ways is healthy for people.''}

Participants also observed that on BeReal they felt less anxiety about others' perceptions and spent less time engaging in social comparison than they do on other platforms. P23 recognized the app's focus on reducing the anxiety generated by metrics such as ``Likes,'' \textit{``I don't really get that many reactions either, but I don't really care, so that's the point of the app. I shouldn't really care for what people are seeing or thinking about me when I'm posting it. That's the point.''} Furthermore, several participants noted that the changes they experienced on the app were self-directed, attributing this to the platform's enjoyable and casual design, as highlighted by one participant who described it as \textit{``fun''} (P06). For example, P04 shared, \textit{``I don't think I've ever felt, like, forced \ldots I feel like the whole idea is kind of, you're not judging people, like, it's not meant to be like that. It's really just for fun, and you're only on there if you want to be on there.''}

\subsubsection{Combatting toxicity with simplicity} BeReal's straightforward design, with its focus on authentic photo sharing, significantly reduces toxicity levels compared to platforms like Snapchat, Instagram, and TikTok. The minimal use of comments on BeReal---often the breeding grounds for toxicity on other platforms---was seen by participants as a key factor contributing to a healthier environment. P20 shared: \textit{``People can be pretty rude on other social platforms. But on BeReal, it's not like Twitter where it's like you can talk to anybody anytime. You have to find the person, you have to friend them. You have to engage on their posts to be able to contact them. It's not as loose as maybe Snapchat or Instagram or Twitter.''}

Many participants also appreciated that, unlike other social media platforms, BeReal's minimalist design does not promote prolonged usage. As P13 shared: \textit{``when I'm looking through it, there's never really anything negative \ldots I don't feel sucked into it \ldots I don't feel I never get on the app and I'm like, oh God I'm going to be on here for another hour or even ten minutes really. I just know it's something I can open and close without like being like, oh one more post, one more video or whatever.''} This view also suggests that BeReal's simplified, chronological algorithm-based feed empowers users with greater control over how and how long they engage with the app's content.

\subsection{The Perceived Downsides of BeReal's Design}
\label{downsides}
Although BeReal has successfully nudged users toward an authenticity norm that mitigates many existing issues associated with mainstream social media platforms, participants raised two shortcomings of BeReal, specifically that it could be both toxic and boring.

\subsubsection{Creating artificial pressure to post} Participants shared that they often feel pressured to be authentic, sometimes even against their will. As P25 shared, \textit{``I definitely feel like I am forced \ldots And so I get curious, and then I end up posting anyways to see what's going on.''} The level of pressure that unfiltered selfies caused varied across participants, where some experienced \textit{``sometimes there are times where I don't have makeup on or I don't look my best and stuff because I was at the gym or something, and so it makes me feel a little stressed out,''} for example. P19 also illustrated this issue, stating, \textit{``You just get so demotivated from feeling like you have to be real, where you're just like, `Well, I don't even want to touch the app at all.'''} Several features, like the lack of filters, retake count, ``late post'' label, public display of reactions, and the implicit pressure to keep up with posting on a daily basis, were identified as stress triggers.

A few participants felt that while the authenticity norm on BeReal carries significance, the application of BeReal's \textit{``rules''} (P27) should prioritize user comfort. As P29 explained: \textit{``I think some people are very obsessed with the idea of keeping things real, keeping things authentic\ldots But to me, I think that it's okay. I think it's okay to be a little bit less authentic online. Only for reasons though, if you have a reason for it, I think that it's okay.''} They believed it acceptable to deviate from BeReal's norms to avoid personal discomfort. P09 explained: \textit{``If I'm self-conscious about it, I won't post about it, and I'll just post about something else.''} Similarly, P16 expressed that when they were not in the mood, they would simply point the camera \textit{``to my ceiling and then one is at the desk.''} The need for positive self-presentation was a particular cause of discomfort. P25 reflected: \textit{``I still have the pressure of trying to make myself look cooler on BeReal. So if I'm just sitting alone, just chilling, watching TikToks, I usually don't post BeReal until I'm out [of] my bed or doing something else at least.''}

\subsubsection{Inciting new forms of toxicity} Despite its intentions, BeReal's simplicity and focus on everyday moments inadvertently fostered new avenues for toxicity, with users sometimes judging each other's content as \textit{``uninteresting''} (P16), \textit{``underwhelming''} (P19), and \textit{``boring''} (P26), either expressing these judgments directly to their friends or confidentially with us during the study interviews. While participants were divided over whether the uninteresting nature of the posts stemmed from the posts themselves or the individuals creating them, some participants perceived that the lack of engaging content was more directly associated with the users. For instance, P27 remarked, \textit{``Some posts show people just lying in bed. I think, `That doesn't seem fun.' If they do this every day, it appears they have no exciting point in their day.''} 

Anticipating or encountering such judgments led some participants to feel an amplified fear of negative judgment when posting on BeReal, which usually involved more vulnerability than posts on other social media platforms in general. As P05 noted, \textit{``People could make fun of your true self.''} This led some participants to \textit{``just walk outside and take a picture''} (P29). The participant further explained, \textit{``Even if they don't necessarily go out, they wait and just leave the house a bit, open the door, take a picture, and then just walk back inside.''} P20 further pointed out the sometimes paradoxical nature of the authenticity norm on BeReal: \textit{``People might judge you negatively because they compare you to the people who aren't authentic.''} Since it is possible that not everyone embraces BeReal's ``authenticity'' to the same extent, being authentic could sometimes lead to negative consequences. P25 shared a candid example: \textit{``I know for me, we all glamorize those super social party raves type of people who hang out, \dots Whereas I feel like when I try to be authentic, it's like I don't do that and usually it's just me at home watching my shows. And so, it definitely conflicts with what society expects versus what I want to do with my life.''}

Another particularly salient issue arises when users begin to critique each other's level of authenticity. P17, for instance, overheard a friend remarking, \textit{``Oh, look, they just retook their BeReal 12 times,''} while P12 shared a similar story about a conversation her sister had over the phone: \textit{``Oh my gosh this person retook their BeReal like several this many times, and I only did it this many times can you believe it?''} This observation was not isolated, and participants P02, P06, P11, and P20 noted similar interactions.

\subsubsection{An underwhelming user experience} BeReal's focus on authenticity ironically leads to a dull and monotonous user experience. As P19 noted, the platform's norms around authenticity, combined with specific posting times and limited features, feels \textit{overbearing.''} Such strict structure and time constraints contribute to repetitiveness and boredom. This renders the overall user experience \textit{underwhelming''} (P19) for some participants. Some users also viewed BeReal's popularity as temporary, sharing \textit{It's one of those fads, in a sense''} (P19), and observing, \textit{I think BeReal kind of died, and it was just a fad''} (P26). BeReal's single-minded focus on authenticity cultivates an \textit{``underwhelming''} user experience that pushes users away from the platform.

\subsection{BeReal's Potential and Limitations to Support Relationship-Building}
Participants also reflected on the extent to which BeReal's encouragement to share unstaged photos at unexpected times might enable relationship-building (or fail to do so). We examined these reflections from the perspective of Social Penetration Theory~\cite{altman1973social} and reported on teens' thoughts about BeReal's potential for increasing the breadth and depth of self-disclosure.

\subsubsection{Enabling casual relationship building} Participants noted how BeReal provides insights into the daily lives of friends, offering a more intimate view of moments they might not otherwise share. P21 gave an example: \textit{``I can see things that I normally wouldn't have known about [my friends]. My friend the other day, I didn't know this, but she got boba at the most random time \ldots it gives me a new insight into what people do in their lives when I'm not with them, so that's why I can appreciate it.''} In this way, BeReal can support relationship building by broadening users' perspectives about others.

Many of our participants also expressed that BeReal is a valuable tool for maintaining connections and staying in touch with their friends. They specifically valued the ability to catch a glimpse of the day-to-day activities of friends who live far away. P04 shared: \textit{``I get to see, like, a little bit of my, some of my friends' days that I wouldn't see otherwise because they don't live in [city name] they live in other places \ldots I wouldn't normally see a picture of them at school or something like that, but it's kind of, it's kind of nice to see \ldots a little glimpse into people's lives.''}

\subsubsection{Limited potential for deepening relationships}
Many of our participants reflected that they want to share different facets of themselves online, suggesting a lack of nuance in the kinds of self-disclosure that BeReal makes possible. They found it limiting that on BeReal they are not able to share \textit{``[their] energetic style or [their] talkative style''} (P28), that they \textit{``like listening to K-pop''} (P09), \textit{``a screenshot\ldots of a game [they're] playing or a song [they're] listening to''} (P16), and \textit{``thoughts of the day or whatever you're doing''} (P14), for example. They wanted to be able to share \textit{``anything from like my camera roll that I took like maybe like ten minutes ago''} (P03), or just \textit{``a lot of things''} (P15) that they would share or see on their social media, Finsta in particular.

Our participants also shared that what they post on BeReal is \textit{``very surface level''} (P15) since it is \textit{``just a photo''} (P05), implying that depth of self-disclosure is also lacking. They expressed that they are not able to share their emotions (P02, P16), personality (P18), values (P27), and political beliefs (P06, P15) as they do on other platforms such as Instagram (P06, P09, P18), Snapchat (P09), or Twitter (P02). They also confirmed that the lack of varied forms of exchange makes it harder for them to learn more about and feel closer to other users on BeReal than on other platforms where deeper self-disclosure happens. P16, for example, shared: \textit{``I think you can't really communicate your emotions that easily through [BeReal], which to me is a pretty important part of getting to know somebody, what kind of things they're going through.''} Such limitations led participants to suggest BeReal adopt other forms of content-sharing, including live photos or text posts.

\subsubsection{An inability to share with specific others} Though BeReal encourages users to curate their audience, it was difficult for some of our participants to draw that boundary given their offline social dynamics. As Social Penetration Theory~\cite{altman1973social} explains, people build relationships by disclosing intentionally within the context of a specific relationship. Users' inability to target who they shared with exacerbated their feeling of intrusion or overwhelmed with making BeReal posts. For example, P17 shared, \textit{``I'm in high school, and so a lot of it is just people in my grade that I wouldn't necessarily trust. \ldots So there have been times where I've retaken my BeReal, because I didn't like the way I looked in a photo. Which I think is not the point or what they originally intended, but I think that in the social situation, like high school is, I feel like that it was bound to happen in a way. And I know people that'll retake their BeReals 10 times just to get the right angle.''} They further emphasized the limits of sharing ``authentic'' content if they are not close with their BeReal friends, stating, \textit{``In my opinion, I think no one is really going to see the authentic you but yourself, or maybe someone you're really close with, like a partner or a really close friend or really close family member. Because everyone around you isn't going to be able to see you truly unless they know you really well.''}

Similarly, some participants perceived unfiltered photo-sharing to be relatively high-stakes and intimate. For example, P25, who called themself a \textit{``homebody,''} expressed that they do not want other people to see that they \textit{``stay at home all the time.''} Such feelings of intrusion were especially evident when participants had friends on BeReal with whom they were not very close. For example, P18 shared, \textit{``I do also worry about what I look like, because there's people that I'm not that close with, that I'm friends with on there, so I care about what I look like, and I cut off my face if I feel like I look bummy.''} 

\subsubsection{Coerced self-disclosure cannot enable relationship building} Some participants explained that for self-disclosure to be authentic, it must be voluntary and comfortable for the individual. P29 expressed, \textit{``the most important thing is that someone has to be comfortable with [sharing]''} and that \textit{``if you're uncomfortable with something, then you're not being real, per se,''} suggesting self-motivation as a prerequisite for authenticity. P19 also shared that BeReal is \textit{``technically authentic, but at a price.''} They elaborated, \textit{``I guess I could call it real, but at the same time, I feel like there's much more just that isn't worth it. You get self-conscious, feel the need to do it, you feel like you have to do it.''}

Our participants emphasized that the authenticity of self-disclosure hinges on the user's intentions. As P19 exemplified, even staged social media photos can be \textit{``perfectly authentic''} if the user consciously acknowledges their staging. Conversely, an `in-the-moment' BeReal with no staging can be considered inauthentic if the disclosure lacks self-motivation or intentionality. Contrary to perceptions equating staging with inauthenticity, participants viewed the intentional sharing of personally significant content as authentic. P29, for example, reflected: \textit{``If you are holding a post and you choose to post later\ldots I think to me [that] is authentic. That shows that this was something important to you\ldots I think it has to mean something to you first, to be authentic.''} Conversely, BeReal posts shared without curation were perceived as lacking the depth of self-disclosure necessary for meaningful social connection.

\subsection{Adolescents' Conclusions about BeReal}
\subsubsection{Bound to fail} Some participants expressed cynicism about the potential for BeReal to achieve its mission of encouraging authentic self-presentation, explaining that the platform cannot escape the fact that on any social media platform, users will seek approval and present overly positive versions of themselves. For example, P17 shared, \textit{``Yes, people are showing themselves kind of unedited, but I think also it was kind of bound to fail in some ways. No matter what, people are going to want to impress other people. No matter what, people are going to want to have other people see them look good.''} They added \textit{``I think any social media platform, no matter what it is, it's never going to be the authentic \ldots because \ldots if you're looking for someone else's approval, that's not being authentic,''} suggesting that authenticity and social media are fundamentally incompatible. P25 shared a related perspective that online self-disclosure should be approached with caution, stating, \textit{``yeah, it is promoting authenticity, but you still have to remind yourself that this is still online, and so your digital footprint is still there, and everything you post will still be there, and people will see that. So don't get too sucked in. Don't glamorize it too much because it's still something that's on the phone and something. It's still considered a social media.''}

Some of our participants also asserted that motivations to preserve private experiences precede self-disclosure on social media platforms. P25 captured this sentiment, saying that displaying everything on social media, including mundane activities like \textit{``enjoying a cup of fruit by yourself''}, can cause users to \textit{``lose a lot of the special aspects''} of their lives. They shared additional insight: \textit{``I feel like if we're too authentic on social media, it takes away the beauty of being connected in real life. If everybody knows what you're doing all the time, then there's no real privacy or connection between the people who see you every day because your every day is already on there too.''} P20 also pointed out that certain aspects of life should be kept private, stating,\textit{``you're not going to post your entire life on social media unless you're a vlogger or an influencer. But even then, you still have parts where you don't want to show other people.''}

\subsubsection{Overly prescriptive} Some participants outright rejected the idea that they should adhere to BeReal's prompts, viewing them as either unnecessary pressure or misaligned with their social media usage. P14 epitomized this perspective, stating that BeReal serves primarily as \textit{``pure entertainment''} and that people use the platform as an \textit{``excuse to take a photo and share it.''} P25 went a step further, expressing clear discomfort with an overly serious approach to the app's rules, stating, \textit{``If someone\ldots need[s] to do a couple retakes within the two minutes, who gives a crap? Even if it's not real, but I think everybody's just taking it too seriously.''} 

Participants reported various ways to circumvent the platform's intended use. For example, they can choose to delay taking photos or retake them multiple times. This flexibility allowed unconvinced users to disregard BeReal's norms entirely, justifying their decision to stage or delay posts for curated self-presentation. P09 illustrated this attitude: \textit{``Personally, I kind of ignore the fact that they tell me to be more real and I kind of just take a photo whenever I feel like}.'' P05 echoed this sentiment, stating: \textit{``I feel the same as like Instagram because, like, you could honestly post whenever you wanted.''}

\subsubsection{Nevertheless, a welcome change} Despite criticisms, the consensus among participants was that BeReal's initiative to foster authenticity is \textit{``a great thing''} (P22)---a positive departure from the norm of other broadcast social media platforms. For example, P06 stated that they appreciated that BeReal \textit{``wants to give authenticity''} to its users because \textit{``it's never been done before and so it's something different to see and \ldots experience.''} P29 also shared that they like the \textit{`` overall message of what [BeReal is] trying to do''} as they feel that BeReal \textit{``encourages people to not fall into social media's expectations for what they should and shouldn't do, or how their life should and shouldn't be.''} P02 went further, suggesting that \textit{``it would be amazing if those apps [i.e., Instagram], or maybe popular people on those apps, would start trends where you were more authentic. So when you got your BeReal, your BeReal notification you would also post something on Instagram to show like vulnerability on all social media apps.''} For those who felt dissatisfied with BeReal's take on ``authenticity,'' they felt that BeReal is \textit{``less inauthentic''} (P10n) at the very least, and that it is \textit{``the most authentic photo version of yourself''} given BeReal photos cannot be edited.

\section{Discussion}
Peer relationships play a crucial role in adolescent development, and reciprocal and authentic self-disclosure is necessary for these relationships to develop. Despite the significant role of social media in facilitating such adolescent peer interactions, its capacity to support authentic self-disclosure is often compromised, for example, by context collapse and the pressures of self-presentation. This study examines BeReal, an app distinguished by its focus on fostering reciprocal, authentic self-presentation.

\begin{figure}[t]
    \centering
    \thispagestyle{empty}
    \includegraphics[width=0.6\linewidth]{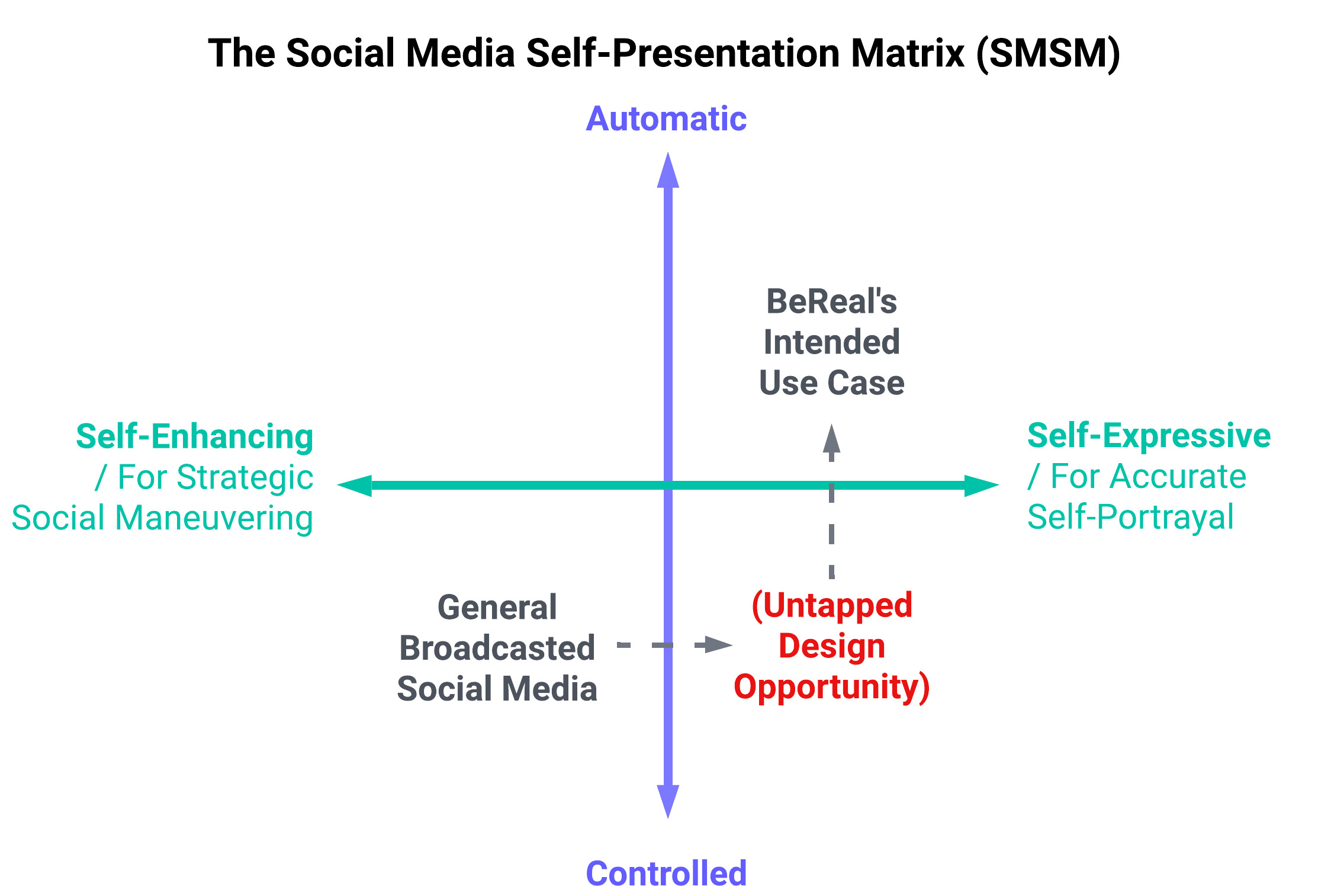}
    \caption{The Social Media Self-Presentation Matrix: A 2x2 matrix with two types of self-presentation goals based on Schlenker's work on self-presentation~\cite{Schlenker2003-yn}. On the horizontal axis: ``Self-Enhancing'' (i.e., strategic social maneuvering goal) and ``Self-Expressive'' (i.e., accurate self-portrayal goal). On the vertical axis, we categorize the cognitive process of self-presentation as either ``Automatic'' (i.e., without conscious awareness, with minimal cognitive effort) or ``Controlled'' (i.e., deliberate, planned). The quadrant of Controlled and Self-Enhancing self-presentation is exemplified by the general strategy of highly manicured self-presentation on broadcast social media platforms like Instagram. BeReal is intended to fall under the Automatic, Self-Expressive quadrant. We identify a design opportunity for social media platforms to move toward the Controlled, Self-Expressive quadrant. This quadrant would represent a more realistic approach toward authenticity, considering the intricacies of social media interactions, such as privacy issues, and the developmental characteristics of teens, such as heightened self-consciousness. We anticipate that a shift toward this quadrant could, in the long run, lead users or platform norms toward Automatic, Self-Expressive self-presentation, which is what BeReal aims to achieve.}
    \Description {The image is a diagram titled ``The Social Media Self-Presentation Matrix (SMSM),'' with two intersecting axes: (1) The vertical axis ranges from ``Automatic'' at the top to ``Controlled'' at the bottom. (2) The horizontal axis ranges from ``Self-Enhancing / For Strategic Social Maneuvering'' on the left to ``Self-Expressive / For Accurate Self-Portrayal'' on the right. Each quadrant represents a different type of self-presentation: 1. Top-left quadrant (Automatic + Self-Enhancing): Not labeled; 2. Top-right quadrant (Automatic + Self-Expressive): Represents self-presentation that is automatic and authentic, aimed at true self-expression. This is where ``BeReal's Intended Use Case'' is located; 3. Bottom-left quadrant (Controlled + Self-Enhancing): Represents self-presentation that is controlled and strategic, typically involving careful curation for the purpose of social image enhancement. This is where ``General Broadcasted Social Media'' is positioned; 4. Bottom-right quadrant (Controlled + Self-Expressive): Represents self-presentation that is controlled and authentic, with an emphasis on accurate self-portrayal. This quadrant is labeled ``(Untapped Design Opportunity)'' in red, suggesting an unexplored area for social media design. Two dashed arrows from ``General Broadcasted Social Media'' to ``(Untapped Design Opportunity),'' and ``(Untapped Design Opportunity)'' to ``BeReal's Intended Use Case'' indicate a two-step path.}
    \label{fig:diagram}
\end{figure}

Our research examines BeReal's design in supporting authentic self-presentation on social media and assesses its alignment with the developmental and self-presentation needs of teens. Here, we (1) introduce the Social Media Self-Presentation Matrix (SMSM, see Figure \ref{fig:diagram}), a conceptual model of the forms that self-presentation can take, (2) discuss the implications of social media design based on the matrix, and (3) propose design recommendations for authentic sharing on social media derived from our study findings.

\subsection{The Social Media Self-Presentation Matrix (SMSM)}

\subsubsection{Evaluating BeReal's Design Support for Self-Presentation Using the SMSM}
Drawing on theories of self-presentation~\cite{Schlenker2003-yn}, we identify two axes for describing self-presentation on social media. The first axis, \textit{Automatic vs. Controlled}, captures the cognitive processes that drive self-presentation. The second axis, \textit{Self-Enhancing vs. Self-Expressive}, captures the goals of self-presentation. Combining these axes, we create a matrix that describes four modes of self-presentation strategies potentially dictating social media platform designs (see Figure \ref{fig:diagram}). We use this matrix to evaluate BeReal's design in supporting authentic self-presentation and suggest alternative ways for achieving BeReal's design goals.

Social media platforms generally promote a Controlled, Self-Enhancing mode of self-presentation, but BeReal has attempted to shift towards a more Automatic, Self-Expressive model. This effort has been partially successful due to several design choices, such as but not limited to: randomly timed notifications that discourage staged sharing, daily ephemeral posting that reduces posting pressure, hidden follower counts that support audience curation, and a minimalist design that underscores the platform's single-minded focus on extemporaneous, unfiltered photo-sharing. These elements collectively foster user behaviors differ from those on platforms like Instagram, and these differences were largely welcomed by users.

However, challenges remain. For example, users have reported feeling new kinds of pressure and discomfort, leading to doubts about the feasibility of authentic self-presentation in social media contexts. Despite BeReal's efforts, some users felt the platform did not support their relationship-building needs, finding it difficult to present themselves accurately through the constrained format of an uncurated, unfiltered photo. 

\subsubsection{Challenges and Considerations for Achieving Automatic, Self-Expressive Self-Presentation on Social Media}
A closer examination of self-presentation theory, as discussed in Section \ref{lab:rw}, may clarify the mismatch between BeReal’s ideal and the reality experienced by teens. BeReal aimed to transition users' general self-presentation strategies from a Controlled, Self-Enhancing model to an Automatic, Self-Expressive one. Achieving this shift requires several conditions, such as a familiar and supportive audience group, minimal discrepancy between one's self-concept and presented self, adequate skills and resources for accurate self-presentation, and low stakes in general, which would then allow for a more Automatic approach.

However, several factors make this shift challenging. Some users have large friend groups that included people they do not fully trust, making Automatic self-presentation unrealistic. Some people feel more self-conscious because the platform encourages them to reveal aspects of their lives they usually prefer to keep private, such as their limited social interactions. Further, BeReal restricts users to a single, unfiltered photo at a randomly assigned time with a two-minute window. This demands significant skill for accurate representation relative to self-presentation on other platforms with no such constraints. Furthermore, the general negativity in the perception of social media exemplified by privacy concerns tends to guide self-presentation behavior in a more Controlled direction rather than an Automatic one.

Most importantly, self-presentation on social media is inherently guided by external validation and cues, demanding a high level of maturity to balance personal and inner concerns with the outer ones effectively---a level of sophistication that may be beyond what can typically be expected of teenagers~\cite{Schlenker2003-yn}. Indeed, we observed instances where BeReal's design features inadvertently prompted adolescents to overlook their internal cues about whether to disclose or withhold, at times pressuring them to share when they would prefer not to.

These observations suggest that transitioning from Controlled, Self-Enhancing self-presentation to Automatic, Self-Expressive self-presentation is not straightforward due to inherent challenges. Social media inherently involves broader social interactions than F2F interactions, meaning that even with a smaller friend group, the audience is larger, making Automatic self-presentation difficult. Additionally, the prevalence of self-enhancement on social media platforms in general makes Self-Enhancing self-presentation the default strategy. However, Automatic self-presentation processes typically emerge in routine situations that are frequently encountered. Therefore, encouraging Self-Expressive self-presentation on social media requires creating opportunities for more Controlled processes to coexist, allowing users to navigate the complex dynamics of online social interactions more deliberately and effectively.

\subsection{Design Recommendations to Support Authentic Self-Presentation Online}
Informed by data derived from our study and the SMSM, we propose design recommendations for fostering authentic self-presentation on social media.

\begin{enumerate} 
    \item \textbf{Support ``accurate'' self-presentation.} As seen in Section \ref{lab:rw}, self-presentation can be driven by truthful motives, such as the desire to ensure others see them as they truly are. However, accurate self-presentation often requires skill and deliberation. Therefore, providing users with the time and opportunities for this is a crucial aspect of authentic self-presentation on social media. For example, platforms could offer daily prompts or missions that encourage authentic reflection and sharing, allowing users to explore the breadth and depth of self-disclosure. Providing more varied modes of expression can additionally facilitate the Automatic aspects of self-presentation, particularly when it enables self-presentation that aligns with consistent self-images and personality traits.

    \item \textbf{Provide air cover for audience management.} Creating a safe space is essential for facilitating Automatic self-presentation. While automatic processes may seem less relevant to authenticity on social media, it is important to recognize that Controlled self-presentations require significant effort and can be mentally exhausting, potentially leading to self-regulation challenges. This exhaustion not only increases problematic social media use but may also diminish users' effectiveness in managing their subsequent self-presentation, such as oversharing. By protecting the cognitive resources of users, particularly children, we can enhance their capacity for more accurate and authentic self-presentation.

    \item \textbf{Scaffold reciprocal sharing.} According to SPT, mutual disclosure is an essential component of social penetration progression. However, lurking behaviors are prevalent on many mainstream social media platforms. To normalize reciprocity and foster deeper social interactions at an individual or community level, designers might consider introducing mechanisms to minimize the vulnerability associated with initiating online engagement. As reported by our participants, daily post notifications acted as a support mechanism for posting and relieved the pressure associated with voluntary sharing on social media. Thus, strategies such as daily reflection prompts might also serve to diminish the vulnerability associated with initiating disclosures and encourage more users to reciprocate self-disclosure.
    
    \item \textbf{Prevent weaponized authenticity.} Participants said that authenticity norms turned toxic when the platform pressured them to over-disclose or enable users to shame one another, police one another, or compete to be the most authentic. Through intentional design decisions, platforms can encourage candid sharing without making an idol of it. For example, a platform might avoid quantifying authenticity by choosing \textit{not} to display the number of times a photo was retaken.
\end{enumerate}

It is important to note that while self-disclosure is generally desirable for social penetration, pursuing it online could introduce potential risks or unforeseen consequences not typically present in offline settings. Younger users, especially those with problematic smartphone use~\cite{Davis2023-zk} are even more likely to disclose beyond what may deemed the appropriate level of self-disclosure. Therefore, when designing to expand the breadth, depth, or frequency of online self-disclosure, it is essential to anticipate potential harmful outcomes to users, such as online harassment or impersonation. Designers should rigorously test their designs with a diverse user group to ensure no targeted group faces heightened risk, implement robust privacy and security measures on the platforms (e.g., disabling screenshots), and possibly guide users to understand and reflect on their desired level of self-disclosure and/or social penetration online.

\subsection{Limitations and Future Work}
Our study has several limitations. The first is that our participant sample includes significantly more people identifying as girls than as boys or non-binary. Further, BeReal introduced many new features since our data collection: (1) the integration of Spotify, where the cover art of the song or podcast a BeReal user is listening to will appear in the bottom right corner of their post, (2) a bonus BeReal where BeReal allows users to post an additional two BeReals throughout the day---but only if they post their first during the allotted two-minute window, (3) branded accounts (e.g., Adidas, Notion), and (4) ``Groups'' which resemble group chats, to name a few. Perceptions of platform norms have likely to have changed with the introduction of the new features. It would be useful for future work to explore how this extra disclosure and extra element of gamification support or undermine healthy and desired forms of authentic self-presentation or create further privacy concerns or pressure.

Future studies might also consider exploring methods to support more comprehensive authentic sharing integral to adolescent relationship-building. Specifically, while BeReal's design indeed offers a new approach to social media---encouraging users to share more authentically---our study's participants highlighted that the platform focuses primarily on one facet of self-disclosure: transparent photo sharing. They also mentioned the presence of potentially harmful, undesired, or incomplete forms of ``authentic'' self-presentation on BeReal. As a result, there remains a need for future research to explore novel designs that could cultivate different modes of sharing that transcend a single form of self-disclosure or self-presentation on social media, moving towards, for example, more intimate self-disclosure or deliberate self-presentation that our participants expressed a desire for.
\section{Conclusion}
In this paper, we explore the impact of design on promoting authenticity in social media. We conducted a qualitative interview study with 29 teen participants who use BeReal, a social media platform created to reduce self-presentation pressures and perceived lack of authenticity that arise from excessive content curation and strategic self-presentation. Our findings demonstrate that the affordances of BeReal effectively generate a space for authentic self-presentation, where users share their unfiltered lives and engage in relatively little social comparison. However, we also find that transparent photo sharing is only a single facet of our teen participants' understanding of authentic self-presentation, and young people are eager for online spaces that allow them to share more than just the details of their day. This study contributes to the important discourse about the influence of social media on teen relationship-building and presents a framework and design guidance for developing platforms that support the different relationship and self-presentation needs of the youth.

\begin{acks}
The authors would like to acknowledge the CERES network, which provided support for this work. We additionally thank the anonymous reviewers for their detailed feedback and the participants for sharing their thoughts. We truly appreciate all the help.
\end{acks}

\bibliographystyle{ACM-Reference-Format}
\bibliography{references}

\clearpage

\appendix

\section{Interview Protocol}
\label{appendix:protocol}

\subsection*{\textless0 Greetings\textgreater (5)}
\begin{enumerate}
    \item Thank you so much for joining me today! Hi, I’m xx, I am a researcher from xx, and I am studying what adolescents and young adults mean by “being real” on social media.
    \item Today we’re going to talk about your experiences with using BeReal, and what you like or dislike about it.
    \item The interview will last no longer than an hour, and we will make sure that none of the information that you share in this interview will be shared with anyone in a personally identifiable way.
    \item Please remember that you should feel free to decline to answer a question if you feel uncomfortable with it.
    \item I wanted to take a moment to ask if you have read the consent agreement and if you have any questions about the consent form?
    \begin{enumerate}
        \item Great. If you have any questions or concerns regarding this interview during or after our interview, please feel free to email me through the contact information on the consent form.
    \end{enumerate}
    \item I don’t work for BeReal, so you don’t need to try to advocate for BeReal if you don’t.
    \item Before we start our interview, do I have your permission to record this interview? The audio may be used in a research paper, but the video will not.
    \begin{enumerate}
        \item Stop interview.
    \end{enumerate}
    \item Great, I’ve just started recording. Today is [date] and I’m talking to [participant ID].
    \item Can you confirm that you have agreed to participate and to be recorded?
\end{enumerate}

\subsection*{\textless1 Rapport / Apprehension\textgreater (10)}
\begin{enumerate}
    \item How long have you been using BeReal, and how often do(did) you use it?
    \item How did you learn about the platform and why did you decide to use it?
    \item How many friends do you think people typically have on BeReal?
    \item How close are you with your BeReal friends?
    \item How did you select the friends that you added on BeReal?
    \item Do you have friends on BeReal that you don’t really trust?
    \item What’s a typical post that you post(ed) on BeReal? Is there an example that you’re okay with sharing?
    \item How would you describe a typical post that your friends post on BeReal?
    \item Which social media platforms, besides BeReal, do you use or have used before? Which ones did you use most frequently?
    \item What are some of the most fulfilling and the most stressful moments that you experience on each platform?
    \item How is BeReal different from Snapchat, Instagram, and TikTok in terms of its vibe?
    \item What are some biggest differences among each community?
    \item How about Instagram and Snapchat?
    \item What are some reasons that people wouldn't post on TikTok Now but would post on BeReal?
\end{enumerate}

\subsection*{\textless2 Exploration\textgreater (10)}
\begin{enumerate}
    \item Do you feel that you’re getting to know your friends better through interactions on BeReal? In what ways?
    \item Which features have you noticed that you think the BeReal creators introduced to encourage people to be more real?
    \item BeReal says that it wants people to be more real online. What’s your impression of that goal? 
    \item Do you think BeReal is doing a good job of making a more authentic social media platform? To what extent?
    \item What are some things they can improve on?
    \item Do you ever feel like you’re being forced to be authentic on BeReal because of all the restrictions?
    \item Do you ever hack their system, for example, by waiting on a notification until you’re ready to post, or do you know that your friends do that?
    \item Is it okay or not okay to bend the rules on BeReal? Why?
    \item What are some moments that you experience toxicity on BeReal, if there are any?
    \item Do you ever feel self-conscious about the things you post on BeReal?
    \item If you do, how do you or do you not overcome that?
    \item Have you ever thought that someone else’s post is boring or that someone else’s life seems boring, seeing what they post on BeReal?
    \item What advice would you give to someone who feels self-conscious on BeReal? For example if they are concerned that people will find their posts to be boring or inappropriate?
    \item Was body image issue ever a concern for you or the people you know on BeReal?
    \item Can you give me metaphors?
    \item Do people seem to compare themselves with other people on BeReal? In terms of \# of likes, \# of friends, what they do on a daily basis, etc.
    \item What’s a cringe post on BeReal? Like that’s weird why would you post that here?
    \item Why do you feel that that is cringy?
    \item Have you heard of any younger/older people that are like 13--15 (16--18) having different views about this? Are they different from the views that friends of your age have?
\end{enumerate}

\subsection*{\textless3 Co-operative\textgreater (15)}
\begin{enumerate}
    \item Have you ever experienced times when you felt differently about other social media platforms because of your experiences on BeReal? 
    \item e.g.) they’re really fake, it feels more wrong to post things there
    \item e.g.) they allow a lot of freedom to post, they are more fun and interesting
    \item Does posting on BeReal feel different from posting on other social media? How?
    \item If there is anything that you would post only on BeReal and not on any other social media, what would it be?
    \item Why is it that on BeReal you can post those things?
    \item What are some things about yourself that are not conveyed well on BeReal?
    \item In your opinion, how would you define authenticity or being real? Why is it important or not important to you?
    \item In your opinion, is authenticity on social media a good thing? Is more authenticity the better?
    \item Are authenticity online and authenticity in daily lives different?
    \item Do you want to experience more authenticity online than what you’re currently experiencing (on BeReal)?
    \item In what context do you feel like you’re the most authentic? At home? On TikTok? At school? With friends?
    \item Can you give an example of a time when you felt like you were being true to yourself?
    \item (if the answer above is offline) What about on online platforms?
    \item Can you think of any backlash or rejection for being authentic on social media?
    \item Do you ever feel any pressure to conform to societal expectations on social media? Does that ever conflict with your efforts to be yourself on social media?
    \item How do you stay authentic to your values and beliefs when faced with peer pressure or temptation?
    \item What’s your impression of the company?
    \item Let’s say you can change current social media in any way possible, from the audience, the design, the features, and everything. What would you like to change the most and why?
    \item When you’re trying to be authentic online, do you find photos to be more effective or text? Temporary vs permanent posts? Larger vs closer groups of friends?
    \item How does BeReal handle privacy and security compared to other platforms you use?
    \item How does BeReal's content moderation compare to other platforms you use?
    \item Does that affect your view of the platform or the company?
    \item Will having advertisements on BeReal affect how you like BeReal?
\end{enumerate}

\subsection*{\textless4 Concluding\textgreater (5)}
\begin{enumerate}
    \item How would you describe your BeReal experience using three adjectives? Why did you choose [words]?
    \item If you could change two things about BeReal, what would they be? The design, the features, the people, anything you can think of.
    \item What would make you stop using BeReal? When would you decide to stop using it?
\end{enumerate}

\end{document}